\documentclass[prd,superscriptaddress,showpacs,preprintnumbers,amsmath,amssymb]{revtex4}
\usepackage{amsmath}
\usepackage{amstext}
\usepackage{amsopn}
\usepackage{amsfonts}
\usepackage{amssymb}
\usepackage{bbm}
\usepackage{accents}
\usepackage{empheq}
\usepackage{graphicx}
\usepackage{epsf}
\usepackage{graphics}
\usepackage[latin1]{inputenc}

\begin{document}
\title{ A note on Low Energy Effective Theory of Chromo Natural Inflation in the light of BICEP2 results}

\author{Anindita Bhattacharjee$^1$, Atri Deshamukhya}
\affiliation{Department of Physics, Assam University, Silchar, Assam 788011, India}

\author{Sudhakar Panda \footnote{on lien from HRI, Allahabad, India}}
\affiliation{Institute of Physics, Bhubaneswar, Odisha 751005, India}

\begin{abstract}
Recent result of BICEP2, revealing a larger value of tensor to scalar ratio ($r$), has opened up new investigations of the inflationary models to fit the experimental data.  The experiment needs to reconfirm the results, specifically the consistency between Planck and BICEP2. On the other hand, the combined analysis of Planck and BICEP2 B, including the dust polarization uncertainty,  brings down the upper limit on $r$.  In this note, we reexamine the low energy effective theory of Chromo Natural Inflation model and its generalization in view of such observational data. We find that the parameter space of the model admits a large value of $r$ as well as other cosmological observables consistent with data.
\end{abstract}


\maketitle

\section{Introduction}
The idea of 'Inflation' was proposed \cite{inflation1,inflation2,inflation3,inflation4,inflation5} to solve a number of cosmological issues associated with standard big bang scenario. However, its real strength was realized with its success in explaining the CMB fluctuations i.e. the large scale structure formation of the universe. The main feature of an inflationary model is a scalar field coupled to gravity slowly rolling down the slope of its potential. To satisfy observational constraints like sufficiently long period of inflation to address the horizon problem and cosmic microwave background limit on density perturbation, the inflaton potential should be extremely flat.  This situation can be achieved in models where the smallness of coupling of inflaton field is protected by a  symmetry of the action since in such a case the quantum correction to the potential in under control. For example, treating the inflaton as a Nambu-Goldstone boson endowed with shift symmetry \cite{Freese}, an inflationary model has been devised in which  the flatness of the potential becomes natural (leading to the term 'Natural inflation'). Natural inflation seemed to be observationally viable  \cite{Freese2}, before BICEP2 results,  but one requires an axionic decay constant of the order of or larger than Planck mass which appears to be difficult to realize in the realm of string theory \cite{BDFG}. Later discovery of viable axionic inflationary models that do not need super-Planckian decay constants (See \cite{decay1,decay2,decay3,decay4, ADS1} for example) led to revival of interest in axionaic inflationary models.

Adshead et al \cite{ADS1} proposed the simplest axionic inflationary model with sub-Planckian axionic decay constant. New ingredient of their model being a collection of non-abelian gauge fields in addition to the axionic field. An SU(2) gauge field is assumed to initiate inflation. This model achieved slow-roll inflation by transfer of axionic energy into classical gauge fields rather than from dissipation via hubble friction. This energy exchange is mediated by the coupling between the axion and Chern-Simon term of the non-abelian gauge field. Since the Chern-Simon term is a total derivative term, it respects the axionic shift symmetry. But this leads to absence of any other axion- gauge field interaction. The higher order corrections to the gauge fields are small and thus the coefficient of the Chern-Simon term could be tuned consistently in a technically natural way to make the potential sufficiently flat. This model is referred as 'Chromo-natural Inflation' in literature. In this model, the structure of gauge coupling generates an additional damping effect for the axion which further slows down its motion \cite{Martinec}. However, it has been observed \cite{EDMP}, \cite{AMW} that when the lenear perturbation of gauge fields are taken in to account, the  theory involving both the gauge field and the axion field develops a perturbative instability. The instability arises due to classical gauge field background which violates parity. Neverthless, this instability disappears when the mass of the vector field fluctuation is much greater than the Hubble scale. On the other hand, Dimastrogiovanni et al \cite{EMA1} demonstrated that when this condition is satisfied, it is possible to integrate out the massive gauge field fluctuations maintaining the modifications introduced by the gauge fields in the dynamics of the axion field. They showed that, in this limit, the chromo-natural inflation is exactly equivalent to a single scalar field effective theory with a non-minimal kinetic term. We will be studying this low energy single field effective theory and its generalization.

The recent results of BICEP2 \cite{BICEP} collaboration, claims an observation of B-mode polarization in the Microwave Background (CMB) whose signal amplitude corresponds to a tensor to scalar ratio $r= 0.2^{+0.07}_{-0.05}$ or $r= 0.16^{+0.06}_{-0.05}$ depending upon the modeling of polarized dust foregrounds though doubts have already been raised on such modeling which reportedly \cite{Morto} could lower the amplitude of the tensor mode signal to the level of becoming undetectable in observations. Single field slow-roll scenarios are seemed to be favoured by BICEP2  from the point of quality of fit and theoretical simplicity \cite{Jerome1}.

In this note, we reinvestigate the  low energy effective theory of Chromo-natural inflation in the light of BICEP2. In section II, we briefly discuss the effective theory and carryout  the analysis for  inflation model   and scan the parameter space  to comment on its viability  in the light of BICEP2 observations. In section III. we consider a simple generalization, namely, the coupling of the axion field with the  Chern-Simon term for the gauge field in the action being a sine function instead of being linear in axion field and investigate its effect in predicting the perturbation spectrum.

\section{Low Energy Effective Theory of Chromo-natural inflation revisited :}

In Chromo-natural inflation, a $SU(2)$-valed gauge field interacts with an axionic scalar field via the Chern-Simon interaction term and the full dynamics is expressed in terms of the following  action containing the gauge field and the pseudo scalar axion field \cite{ADS1},\cite{EMA1}:
\begin{equation}
S_{chromo}=\int~d^{4}x~\sqrt{-g}~\left[-\frac{M_{P}^{2}}{2} R- \frac{1}{4} F_{\mu \nu}^{a} F_{a}^{\mu \nu}-\frac{1}{2} (\partial \chi)^{2}
-\mu^{4} \left( 1+\cos \left( \frac{\chi}{f} \right) \right) +\frac{\lambda}{8 f} \chi \epsilon^{\mu \nu \rho \sigma} F_{\mu \nu}^{a} F_{\rho \sigma}^{a} \right]
\end{equation}
where  $F_{\mu \nu}^{a}= \partial_{\mu} A_{\nu}^a -\partial_{\nu} A_{\mu}^a- \bar{g} f^{a b c} A_{\mu}^b A_{\nu}^c $,  $\chi$ and $A_{\mu}^a$ are the axion and the $SU(2)$ gauge fields respectively. Here, $\bar{g}$ is the gauge coupling constant, $f$ is the axion decay constant, $\mu$ is the mass scale of the theory, $\lambda$ is the gauge parameter  and  $f^{a b c}$ is the structure constant  of the gauge group with normalization condition $f^{1 2 3}=1$ .  $\epsilon^{\mu \nu\rho \sigma}$ is chosen as $\epsilon^{0 1 2 3}= \frac{1}{\sqrt{-g}}$ with standard FRW metric $ds^{2}= -N^2 dt^2 + a^2(t)\delta_{i j} dx^i dx^j$, $N$ being the Laps function . The last term in the action is the Chern-Simon term which describes the interaction between the gauge and the axionic field. The inclusion of this term in the action gives rise to an additional damping effect which makes this model different from the other inflationary models involving axion field.

Using the $SU(2)$ gauge freedom which allows for an isotropic configuration for $A_i^a = \delta_i^a a(t) \psi (t)$ and $A_0^a = 0$, where $a(t)$ is the scale factor, one can rewrite  the preceding action as an action involving two scalar fields ($\chi$ and $\psi$) coupled to gravity \cite{EMA1}. As mention in the previous section, the linear fluctuation of the gauge field introduces a perturbative instability which disappear when the mass of the fluctuation is much heavier  than the Hubble scale. Further, it has been established in \cite{EMA1} that precisely in this condition one can integrate out the  gauge field resulting in  a single field effective action involving only the axion field. The  new  effective action takes  the form :
\begin{equation}
S = \int~d^{4}x~\sqrt{-g}~\left[-\frac{M_{P}^{2}}{2} R- \frac{1}{2} \left(\partial \chi \right)^{2}+ \frac{1}{4 \Lambda^{4}}\left(\partial \chi \right)^{4} - \mu^{4} \left( 1+\cos \left(\frac{\chi}{f} \right) \right) \right].
\end{equation}
Note that the effect of the gauge field in the single field effective theory is captured by the term $\frac{1}{4 \Lambda^{4}}\left(\partial \chi \right)^{4}$ where $\Lambda^{4}=\frac{8 f^{4} {\bar{g}}^{2}}{\lambda^{4}}$. This important observation is the key factor which distinguishes chromo-natural inflation from other axion inflationary models.

The above effective action is of the form :
\begin{equation}
S = \int d^{4}x \sqrt{-g} \left[ -\frac{M_{P}^{2}R}{2} + P(X,\chi)\right],
\end{equation}
with $X=-\frac{1}{2}(\partial\chi)^2$ and the matter Lagrangian density
\begin{equation}
P(X,\chi)= X + \frac{X^{2}}{\Lambda^{4}}- V(\chi)
\end{equation}
where, $V(\chi) =\mu^{4} \left( 1+\cos \left( \frac{\chi}{f} \right) \right)$ is the potential of the axionic field $\chi$.

Now, the basic dynamical equations defining the inflationary scenario i.e, the Friedmann equation and the equation of motion of the inflaton field are \cite{MUK1,MUK2,MUK3,MUK4,MUK5,MUK6}
\begin{eqnarray}
H^2~&=&~\frac{\rho}{3 M_P^2},\\
\dot{\rho}~&=&~-3H \left( \rho+P\right),
\end{eqnarray}
where, $\rho= 2X P_{,X}-P$ is the energy density of the scalar field, $P$ is the pressure term  which is same as the matter Lagrangian density in this case and  $(P_{,X} )$ denotes derivative of $P$ with respect to $X$.

It may be noted that in the present context , the speed of sound  $c_{s}\neq 1$ \cite{BABI,BAB2} but can be expressed as
\begin{equation}
c_{s}^{2}\equiv \frac{\partial \rho}{\partial P}=\frac{P_{,X}}{P_{,X}+ 2X P_{,XX}}.
\end{equation}

Also, in the slow roll regime, one can express the  the slow-roll parameters $\varepsilon$ and $\eta$ in terms of $X$ as
\begin{eqnarray}
\varepsilon &\equiv & -\frac{\dot{H}}{H^{2}}= \frac{X P_{,X}}{M_{P}^{2} H^{2}}, \\
\eta &\equiv & -\frac{\dot{\varepsilon}}{\varepsilon H}=\frac{6 X P_{,X}}{2 X P_{,X}-P}.
\end{eqnarray}
During the inflationary phase, the slow roll parameters have to satisfy the conditions $\varepsilon \ll 1$, $\mid\eta\mid \ll 1$ . The end of inflation is marked by the inflaton field value at which either of the two parameters becomes of the order of  unity (whichever is earlier).

 The cosmological perturbation analysis for the model described by the equations (3) and (4) yields the power spectrum for the scalar fluctuations as :
\begin{equation}
P_{k}^{\zeta}= \frac{1}{36 \pi^{2} M_{P}^{4} c_{s} \left( \rho+ P \right)}=\frac{H^{2}}{8 \pi^{2} M_{P}^{2} c_{s} \varepsilon}
\end{equation}
which is evaluated at the time of horizon exit, i.e, at $c_{s} K= a H$ \cite{BHATT1,BHATT2}. Similarly, the power spectrum for the tensor perturbation can be expressed as
\begin{equation}
P_{k}^{h}= \frac{2 \rho}{3\pi^{2}M_{P}^{4}}.
\end{equation}
In terms of these scalar and tensor power spectrum, we can define the three important cosmological observables, namely, the spectral index $n_{s}$,  the running spectral index $\alpha_{s}$ and tensor to scalar ratio $r$ as
\begin{eqnarray}
n_{s} - 1 &=& \frac{d \ln P_{k}^{\zeta}}{d \ln k}, \\ \nonumber
\alpha_{s}&=& \frac{d n_{s}}{d \ln k},\\ \nonumber
r&=& \frac{P_{k}^{h}}{P_{k}^{\zeta}}.
\end{eqnarray}

In the model defined by equations (3) and (4), the energy density is
\begin{equation}
\rho=X+ \frac{3 X^{2}}{\Lambda^{4}}+V(\chi)\nonumber.
\end{equation}
Using the slow-roll conditions which essentially demand $\rho\simeq V(\chi)$ in the slow roll regime i.e, the maximum energy contribution comes from the potential of the axion field,  the Friedmann equation and the equation of motion of the axion field take the forms :
\begin{eqnarray}
H^2~&=&~\frac{V(\chi)}{3 M_P^2},\\
3H \dot{\chi}(1+\frac{\dot{\chi}^{2}}{\Lambda^{4}})~&=&~ \frac{\mu^{4}}{f} \sin \left(\frac{\chi}{f} \right).
\end{eqnarray}

In Ref.\cite{EMA1}, the inflationary scenario has been investigated assuming that the cubic term in $\dot{\chi}$ of equation (14)  dominates over the linear term and hence the linear term has been neglected. With such an  assumption, the major observables i.e the spectral index $n_{s}$ and tensor-scalar ratio $r$ were reported to be in the range [0.951, 0.975] and [0.01, 0.03] respectively for $\frac{\chi}{f}$ in the range [0.1, 2.3]. The authors, however, did not  report  the running of spectral index  in the chosen range of field variable and other model parameters. Note that the reported range of $r$ is consistent with the observation of Planck but is  disfavoured by the recent BICEP2  results.

We re-examine the model by retaining both the linear and cubic terms in the equation of motion for the axion field. This modifies the solution for $\dot{\chi}$ which in turn modifies the dependence of observables on the model parameters.

Let us recall here that the slow-roll hierarchy scaling (SRHS)  demands the condition $X \gg\Lambda^4$ \cite{EMA1} and with this condition, the expression for the slow-roll parameters $\varepsilon$, $\eta$ and the speed of sound $c_{s}$ becomes :
\begin{eqnarray}
\varepsilon~&=&~ \frac{2 X^2}{H^{2} M_{P}^{2} \Lambda^{4}}, \\
\eta~&=&~ \frac{12 X^2}{\left( 3 X^2+ \Lambda^4 V(\chi)\right)}, \\
c_{s}^{2}~&=&~\frac{1}{3}.
\end{eqnarray}
Also, in this limit, the power spectra for scalar and tensor perturbations take the form :
\begin{eqnarray}
P_{k}^{\zeta} ~&=&~ \frac{\sqrt{3} \mu^{8}(1+\cos \left(\frac{\chi}{f} \right))^{2} \Lambda^{4}}{36 \pi^{2} M_{P}^{4} (\dot{\chi})^{4}}, \\
P_{k}^{h}~&=&~ \frac{2 \mu^{4}}{3 \pi^{2} M_{P}^{4}} (1+\cos \left(\frac{\chi}{f} \right)).
\end{eqnarray}
The number of e-foldings when the inflaton field $\chi$ rolls from its value $\chi_{i}$ to $\chi_{f}$ is
\begin{eqnarray}
N_e \equiv \int H dt= \int_{\chi_{i}}^{\chi_{f}} \frac{H}{\dot{\chi}} d\chi.
\end{eqnarray}

For further analysis, we solve equation (14) for $\dot{\chi}$. Out of the three solutions, two  turn out to be complex which we ignore and  retain the only real solution to evaluate all the observables. We do not present it explicitly here as the expression is a bit lengthy.  We carry out the numerical analysis and report our result in the following tables for different number of e-foldings ($N_e = 40, 50$ and $60$) before the end of inflation. We emphasize that $N_e$ denote the number of e-foldingd before end of inflation where the observables are computed and not the total number of e-foldings allowed in the model which plays a lesser role. Here, the inflation  ends  when $\epsilon$ reaches unity ( as it  tends to one faster than $\eta$ in this case). This fixes $\chi_f$ and the value of $N_e$ fixes $\chi_i = \chi_*$ in terms  of the three model parameters  $\mu, \Lambda$ and $f$ at which all the observables are computed.  We  then scan the parameter space to find regions where the spectral index lie within the PLANCK bound \cite{Ade} after satisfying COBE normalization condition $P_{k}^{\zeta}= 2\times 10^{-9}$.  The values of model parameters, obtained in this way, are used to evaluate other observables.  We present below our results in tables I. II and III for $N_e = 40, 50$ and $60$ respectively.
\vskip .5cm

\begin{table}[htbp]
\renewcommand{\arraystretch}{1.6}
\caption{Possible values of the observables for different values of model parameters $f$,$\mu$ and $\Lambda$ \\ (in units of $M_{P}$) for Chern-Simon term of the form $\frac{\lambda}{8 f} \chi \epsilon^{\mu \nu \rho \sigma} F_{\mu \nu}^{a} F_{\rho \sigma}^{a}$ with  $N=40$: }
\label{tab:example}
\centering
\begin{tabular}{|c|c|c|c|c|c|} \hline
    $\mu$   &   $f$   &   $\Lambda$   &  $n_{s}$   &    $r$   &   $\alpha_{s}$  \\
\hline
    0.01   &   0.01   &   0.000001   &   0.950553   &   0.110265   &   -0.0012196 \\
\hline
    0.01   &   0.1   &   0.00001   &   0.950554   &   0.110261   &   -0.00121958 \\
\hline
    0.01   &   0.5   &   0.00005   &   0.95054   &   0.110225   &   -0.00122025 \\
\hline
    0.01   &   1   &   0.0001   &   0.950538   &   0.110016   &   -0.00122026 \\
\hline
    0.01   &   5   &   0.0005   &   0.950477   &   0.103476   &   -0.00122068 \\
\hline
    0.01   &   9   &   0.0009   &  0.950254   &   0.0894433   &    -0.000915847  \\
\hline
\end{tabular}
\end{table}

\begin{table}[htbp]
\renewcommand{\arraystretch}{1.6}
\begin{center}
\caption{Possible values of the observables for different values of model parameters $f$,$\mu$ and $\Lambda$ \\ (in units of $M_{P}$) for Chern-Simon term of the form $\frac{\lambda}{8 f} \chi \epsilon^{\mu \nu \rho \sigma} F_{\mu \nu}^{a} F_{\rho \sigma}^{a}$ with $N_e=50$ :}
\end{center}
\label{tab:example}
\centering
\begin{tabular}{|c|c|c|c|c|c|} \hline
    $\mu$   &   $f$   &   $\Lambda$   &  $n_{s}$   &    $r$   &   $\alpha_{s}$  \\
\hline
    0.01   &   0.01   &   0.000001   &   0.96033   &   0.0876458   &   -0.000783796 \\
\hline
    0.01   &   0.1   &   0.00001   &   0.960333   &   0.0876428   &   -0.000783783 \\
\hline
    0.01   &   0.5   &   0.00005   &   0.960324   &   0.0876096   &   -0.000784134 \\
\hline
    0.01   &   1   &   0.0001   &   0.96032   &   0.0874465   &   -0.000784212 \\
\hline
    0.01   &   5   &   0.0005   &   0.960267   &   0.0822008   &   -0.00078424 \\
\hline
    0.01   &   10   &   0.0007   &  0.960228   &   0.076007   &    -0.000786553  \\
\hline
\end{tabular}
\end{table}

\begin{table}[htbp]
\renewcommand{\arraystretch}{1.6}
\caption{Possible values of the observables for different values of model parameters $f$,$\mu$ and $\Lambda$ \\ (in units of $M_{P}$) for Chern-Simon term of the form $\frac{\lambda}{8 f} \chi \epsilon^{\mu \nu \rho \sigma} F_{\mu \nu}^{a} F_{\rho \sigma}^{a}$ with $N_e=60$ :}
\label{tab:example2}
\centering
\begin{tabular}{|c|c|c|c|c|c|} \hline
    $\mu$   &   $f$   &   $\Lambda$   &  $n_{s}$   &    $r$   &   $\alpha_{s}$  \\
\hline
    0.01   &   0.01   &   0.000001   &   0.966876   &   0.0725007   &   -0.000545603 \\
\hline
    0.01   &   0.1   &   0.00001   &   0.966872   &   0.0725081   &   -0.000545736 \\
\hline
    0.01   &   0.5   &   0.00005   &   0.966874   &   0.0724578   &   -0.000545655 \\
\hline
    0.01   &   1   &   0.0001   &   0.966864   &   0.0723371   &   -0.000545925 \\
\hline
    0.01   &   5   &   0.0005   &   0.966811   &   0.0674989   &   -0.000545901 \\
\hline
    0.01   &   10   &   0.0007   &  0.966788   &   0.0630496   &    -0.000547662  \\
\hline
\end{tabular}
\end{table}


We note that, in this model, for COBE normalization to be respected and to have at least 40 e-folds, we do not  have much freedom to choose $\mu$ and $\Lambda$. Though small changes are allowed around reported values, the order of magnitude seems to have no flexibility. We decided to avoid reporting such allowed small changes in $\mu$ and kept it fixed at a representative value . It is observed, from the above tables, that we can find a set of model parameters for which we can have minimum number of 40 e-foldings and we can satisfy both COBE normalization and the value of spectral index consistent  with observation.  For the same set of parameters, the value of $r$ seems to increase as $N$ decreases but it never reaches the lower bound of the value predicted by BICEP2 i.e $r=0.15$.  However, we see that the inclusion of the linear term in the equation of motion for the axion field results in increase of the value of $r$ from a maximum of $0.03$ as reported in Ref.\cite{EMA1} to $0.11$. We also note that the running of spectral index is one order less than the value predicted by PLANCK.

In the next section, we generalize the Chern-Simon coupling and re-examine the model aiming at the possibility  of  obtaining higher values of the tensor to scalar ratio ($r$) as predicted by BICEP2.

\section{Chromo-natural inflation with generalized Chern-Simon term :}

In the last section, we considered the linear coupling of the axion field with a Chern-Simon term. This is the standard form used in the literature. The Chern-Simon term being odd under parity, one can, in principle, consider  any parity-odd function of the axion field coupled to it. The linear term can be thought of as the minimal coupling. In the context of particle physics,  Pecci-Quin symmetry prefers the linear coupling. However, in the cosmological context such a constraint  is not mandatory and we can consider generalizing it to any odd function of the axion field. For simplicity, we consider the function to be a sine function (i.e. $\sin \left(\frac{\chi}{f} \right)$ is taken as the Chern-Simon term coupling instead of $\frac{\chi}{f}$ )  to investigate the inflationary dynamics of axion and gauge field  and write the action  as :
\begin{equation}
S_{chromo}=\int~d^{4}x~\sqrt{-g}~\left[-\frac{M_{P}^{2}}{2} R- \frac{1}{4} F_{\mu \nu}^{a} F_{a}^{\mu \nu}-\frac{1}{2} (\partial \chi)^{2}
-\mu^{4} \left[ 1+\cos \left( \frac{\chi}{f} \right) \right] +\frac{\lambda}{8} \sin\left(\frac{\chi}{f}\right) \epsilon^{\mu \nu \rho \sigma} F_{\mu \nu}^{a} F_{\rho \sigma}^{a} \right].
\end{equation}
In the above action all the quantities are defined in the same way as in section II. We note that the Chern-Simon term, in the above, explicitly breaks the shift symmetry  which is not an issue in the present context since the potential term also breaks the shift symmetry.

Now, using the $SU(2)$ gauge freedom, one can write the temporal and the spatial components of the gauge field as in the previous section. The above action then takes the form:

\begin{equation}
S_{chromo}=\int~d^{4}x~N a^3 \left[ -3 M_{P}^2 \frac{\dot{a}^2}{N^2 a^2}+ \frac{\dot{\chi}^2}{2 N^2} - \mu^{4} \left[ 1+ \cos \left( \frac{\chi}{f} \right) \right] + \frac{3}{2 N^2 a^2} \left(\frac{\partial(\psi a)}{\partial t} \right)^2 - \frac{3}{2} \bar{g}^2 \psi^4  + \frac{\lambda \bar{g}}{f N} \cos \left( \frac{\chi}{f} \right) \psi^3 \dot{\chi} \right]
\end{equation}

The Friedmann equation which is obtained by varying the above  action with respect to the Laps function $N$ is not effected by the generalization of the Chern-Simon coupling and hence remains same as in the previous section i.e as in Ref.\cite{EMA1}.However, the new coupling modifies  the equation of motion for the scalar fields  and this modification of the field dynamics affects the cosmological observables.

For further analysis, we don't proceed with this two field theory. Rather, following Ref.\cite{EMA1}, we transform the action given by equation (21) into an effective single field action.

\subsection{Transformation into a single field effective theory :}

For this  purpose, it is  convenient to scale the gauge fields and the coupling $\lambda$ as
\begin{equation}
 A_{\mu}^a= \frac{1}{\sqrt{\bar{g}}}\tilde{A_{\mu}^a},~ \lambda=\sqrt{\bar{g}}\tilde{\lambda}.
\end{equation}
In the SRHS limit, i.e when the mass  of the gauge field  is  considered much bigger than the Hubble parameter H (in other words implementing  the condition $\bar{g} \rightarrow \infty$ , keeping $\tilde{A_{\mu}^a}$ and $\tilde{\lambda}$ fixed), the action (21) takes the form :
\begin{eqnarray}
S_{SRHS}&=&\int~d^{4}x~\sqrt{-g}~[-\frac{M_{P}^{2}}{2} R- \frac{1}{4}f^{a b c} \tilde{A}_{\mu}^{b}\tilde{A}_{\nu}^{c}f^{a d e}\tilde{A}_{d}^{\mu}\tilde{A}_{e}^{\nu}
-\frac{1}{2} (\partial \chi)^{2}
-\mu^{4} \left[ 1+\cos \left( \frac{\chi}{f} \right)\right] \\
&-& \frac{\tilde{\lambda}}{2} \sin(\frac{\chi}{f}) \epsilon^{\mu \nu \rho \sigma} \partial_{\mu}\tilde{A}_{\nu}^{a} f^{a b c} \tilde{A}_{\rho}^{b} \tilde{A}_{\sigma}^{c}]\\
&=& \int~d^{4}x~\sqrt{-g}~[-\frac{M_{P}^{2}}{2} R- \frac{1}{4}\left(\tilde{A}_{\mu}^{a} \tilde{A}_{a}^{\mu}\right)^2+ \frac{1}{4}\left(\tilde{A}_{\mu}^{a} \tilde{A}_{\nu}^{a}\right)\left(\tilde{A}_{b}^{\mu} \tilde{A}_{b}^{\nu}\right)
-\frac{1}{2} (\partial \chi)^{2}-\mu^{4} \left[ 1+\cos \left( \frac{\chi}{f} \right) \right] \\
&+& \frac{\tilde{\lambda}}{6}\partial_{\mu} \left[\sin \left(\frac{\chi}{f}\right)\right] \epsilon^{\mu \nu \rho \sigma} f^{a b c} \tilde{A}_{\nu}^{a} \tilde{A}_{\rho}^{b} \tilde{A}_{\sigma}^{c}]
\end{eqnarray}

For the last term in the above equation we have carried out a partial integration. As observed in Ref.\cite{EMA1}, in  the above  action, gauge fields only appear algebraically. As the action is quadratic in $\tilde{A}_{0}^{a}$ , the resulting equation of motion can be easily solved  for $\tilde{A}_{0}^{a}$. But the solution for $\tilde{A}_{i}^{a}$ can not easily be found out as the action is not quadratic in $\tilde{A}_{i}^{a}$, rather have a general algebraic form.  Therefore, it is solved at a single space time point,  $x_{p}$, in a locally inertial rest frame. In fact, we can write  a time-like four-vector $\partial_{\mu}\left[\sin \left(\frac{\chi}{f}\right)\right] $ at the point $x_p$ as a Lorentz boost of a vector pointing only in the time direction and having the same  invariant form as follows:
\begin{equation}
\partial_{\mu}[\sin \left(\frac{\chi}{f}\right)](x_{p})= \Lambda_{\mu}^{\nu}(x_{p})\delta_{\nu 0}\sqrt{-\left[\partial[\sin \left(\frac{\chi}{f}\right)](x_{p})\right]^2},
\end{equation}
where, $\Lambda_{a}^{A}(x_{p}) \Lambda_{b}^{B}(x_{p}) \eta_{A B}= \eta_{a b}$. Similarly, for the gauge field we can write
\begin{equation}
\tilde{A}_{\mu}^a(x_{p})= \Lambda_{\mu}^{\nu}(x_{p})\bar{A}_{\nu}^a(x_{p}).
\end{equation}

Local Lorentz invariance of the action can be used to write  the gauge-field part of the  Lagrangian as :
 \begin{eqnarray}
 \textit{L}_{A}(x_{p})&=& - \frac{1}{4}\left(\bar{A}_{\mu}^{a} \bar{A}_{a}^{\mu}\right)^2+\frac{1}{4}\left(\bar{A}_{\mu}^{a} \bar{A}_{\nu}^{a}\right)\left(\bar{A}_{b}^{\mu} \bar{A}_{b}^{\nu}\right)+ \frac{\bar{\lambda}}{6}\sqrt{-(\partial (\sin (\frac{\chi}{f}))(x_{p}))^2 } \delta_{\nu 0}\epsilon^{\mu \nu \rho \sigma} f^{a b c} \bar{A}_{\mu}^{a} \bar{A}_{\rho}^{b} \bar{A}_{\sigma}^{c}\\
 &=& - \frac{1}{2}\left(\bar{A}_{0}^{a}\right)^2 \left(\bar{A}_{i}^{b}\right)^2 + \frac{1}{2}\left(\bar{A}_{0}^{a} \bar{A}_{i}^{a}\right)\left(\bar{A}_{b}^{0} \bar{A}_{b}^{i}\right)-\frac{1}{4}\left(\bar{A}_{i}^{a} \bar{A}_{a}^{i}\right)^2+ \frac{1}{4}\left(\bar{A}_{i}^{a} \bar{A}_{j}^{a}\right)\left(\bar{A}_{b}^{i} \bar{A}_{b}^{j}\right)\\
 &+&\frac{\tilde{\lambda}}{6}\sqrt{-(\partial (\sin (\frac{\chi}{f}))(x_{p}))^2}\epsilon^{i j k} f^{a b c} \bar{A}_{i}^{a} \bar{A}_{j}^{b} \bar{A}_{k}^{c}.
\end{eqnarray}
Now, varying the action with respect to $\bar{A}_{0}^{a}$ and remembering the fact that $\bar{A}_{i}^{a}$ is having non-zero value and the action is quadratic in $\bar{A}_{0}^{a}$, one gets
\begin{equation}
\bar{A}_{0}^{a}(x_{p})=0.
\end{equation}
With this value, $\textit{L}_{A}(x_{p})$ becomes:
\begin{equation}
\textit{L}_{A}(x_{p})=-\frac{1}{4}\left(\bar{A}_{i}^{a} \bar{A}_{a}^{i}\right)^2+ \frac{1}{4}\left(\bar{A}_{i}^{a} \bar{A}_{j}^{a}\right)\left(\bar{A}_{b}^{i} \bar{A}_{b}^{j}\right)+\frac{\bar{\lambda}}{6}\sqrt{-(\partial (\sin (\frac{\chi}{f}))(x_{p}))^2}\epsilon^{i j k} f^{a b c} \bar{A}_{i}^{a} \bar{A}_{j}^{b} \bar{A}_{k}^{c}.
\end{equation}
This yields a cubic equation after variation with respect to $\bar{A}_{i}^{a}$ which can have three different solutions. It is found that  $\bar{A}_{i}^{a}= \delta_{i}^{a} \frac{\bar{\lambda}}{2} \sqrt{-(\partial (\sin (\frac{\chi}{f}))(x_{p}))^2}$ satisfies the cubic equation and can be taken as the background solution of the chromo natural inflation. It is interesting to note that this solution is of similar form as was originally chosen as the isotropic background solution.

Putting back this in equation (34), we get
\begin{equation}
\textit{L}_{A}(x_{p})= \frac{1}{4 \bar{\Lambda}^4}\left[\partial (\sin (\frac{\chi}{f}))(x_{p})  \right]^4,
\end{equation}
where, $\bar{\Lambda}^4= \frac{8 \bar{g}^2}{\lambda^4}$. As this is evaluated in a local inertial frame, using the principle of covariance, one can take the general form of the Lagrangian as $\textit{L} = \frac{1}{4 \bar{\Lambda}^4}\left[\partial (\sin (\frac{\chi}{f}))\right]^4$.

Finally, including this part, the effective action becomes
\begin{equation}
S_{\chi}=\int~d^{4}x~\sqrt{-g}~\left[-\frac{M_{P}^{2}}{2} R- \frac{1}{2} \left(\partial \chi \right)^{2}+ \frac{1}{4 \bar{\Lambda}^{4} f^4} \cos^{4} \left(\frac{\chi}{f}  \right)\left(\partial \chi \right)^{4} - \mu^{4} \left( 1+\cos \left(\frac{\chi}{f} \right) \right) \right].
\end{equation}

Thus, it is evident that, within the SRHS limit, the chromo inflationary scenario with a more general form of Chern-Simon term can be transformed in to a  single field effective theory where, the effect of the gauge field is captured by the non-minimal kinetic term for the axionic scalar field in the single field action.

\subsection{Chromo inflationary dynamics in single field effective theory }

It is clear from equation (36) that this single field action for the chromo inflation is a single field  theory with the matter Lagrangian density  $P(X,\chi)$ given by
\begin{equation}
P(X,\chi)= X + \frac{\cos^{4}\left(\frac{\chi}{f} \right) X^{2}}{\Lambda^{4}}- V(\chi)
\end{equation}
where, $\Lambda ^4= \bar{\Lambda}^4 f^4 = \frac{8 f^{4} {\bar{g}}^{2}}{\lambda^{4}}$ and $X$ is same as in section II. The form of the energy density is found to be
\begin{equation}
\rho=X+\frac{3 \cos^{4}\left(\frac{\chi}{f} \right) X^{2}}{\Lambda^{4}}+V(\chi).
\end{equation}

Now, from this action, the Friedmann equation and the background equation of motion of the axionic scalar field can be expressed as :
\begin{eqnarray}
H^2&=&\frac{\rho(X,\chi)}{3 M_P^2},\\
\ddot{\chi}\left(1+\frac{3}{\Lambda^4} \cos^4 (\frac{\chi}{f} ){\dot{\chi}}^2\right)&=&-3H \dot{\chi}\left(1+\frac{1}{\Lambda^4}\cos^4 (\frac{\chi}{f} )\dot{\chi}^2\right)
+\frac{\mu^{4}}{f} \sin (\frac{\chi}{f} )
\end{eqnarray}
where we have neglected a term proportional to square of the slow-roll parameter. The slow-roll condition demands that $\rho\simeq V(\chi)$ ie, the maximum energy contribution comes from the potential of the axion and as in this regime, $\ddot{\chi}$ is negligible in comparison to others and hence also can be neglected . Thus, the above two basic equations for this model take the forms :
\begin{eqnarray}
H^2~&=&~\frac{V(\chi)}{3 M_P^2},\\
3H \dot{\chi}\left(1+ \frac{1}{\Lambda^4} {\cos^4 (\frac{\chi}{f}){\dot\chi}^2} \right)~&=&~ \frac{\mu^{4}}{f} \sin (\frac{\chi}{f} ).
\end{eqnarray}

From the cubic equation (42), one can obtain the expression for $\dot{\chi}$ and use this to evaluate all the observables in terms of parameters of this model. Again, this being a long expression, we refrain from displaying it here. Now, in the slow-roll hierarchy scaling limit ie, $X \cos^4(\chi/f))\gg\Lambda^4$, the expression for the slow-roll parameters $\varepsilon$, $\eta$ and the speed of sound $c_{s}$ are determined to be:
\begin{eqnarray}
\varepsilon~&=&~ \frac{2 \cos^4 (\frac{\chi}{f} ) X^2}{H^{2} M_{P}^{2} \Lambda^{4}}, \\
\eta~&=&~ \frac{12 \cos^4 (\frac{\chi}{f} ) X^2}{3 \cos^4 (\frac{\chi}{f} ) X^2+ \Lambda^4 V(\chi)}, \\
c_{s}^{2}~&=&~\frac{1}{3}.
\end{eqnarray}

Following the previous section, the power spectrum for the scalar and the tensor perturbations can be expressed respectively as
\begin{eqnarray}
P_{k}^{\zeta} ~&=&~ \frac{\sqrt{3} \mu^{8}(1+\cos (\frac{\chi}{f} ))^{2} \Lambda^{4}}{36 \pi^{2} M_{P}^{4} (\dot{\chi})^{4}\cos^4 (\frac{\chi}{f} )}, \\
P_{k}^{h}~&=&~ \frac{2 \mu^{4}}{3 \pi^{2} M_{P}^{4}} (1+\cos (\frac{\chi}{f} )).
\end{eqnarray}
From the above expressions of the power spectrum $P_{k}^{\zeta}$ and $P_{k}^{h}$, we can derive the observables $n_{s}$, $\alpha_{s}$ and $r$  as defined in section II ( equation(12)). and they can be computed for different sets of  model parameters. In the following tables we display these values for  some choice of  parameters $\mu, \Lambda$ and $f$ so that the power spectrum satisfies the COBE normalization constraint ie, $P_{k}^{\zeta}\sim 2\times 10^{-9}$. It may also be noted than , in this case the end of inflation is marked by $\eta=1$ which runs faster than $\varepsilon$, contrary to the previous section.

\begin{table}[htbp]
\renewcommand{\arraystretch}{1.6}
\caption{Different range of model parameters  $f$,$\mu$ and $\Lambda$ (in units of $M_{P}$) with Chern-Simon term of the form $\frac{\lambda}{8} \sin\left(\frac{\chi}{f}\right) \epsilon^{\mu \nu \rho \sigma} F_{\mu \nu}^{a} F_{\rho \sigma}^{a} $ when $N=40$:}
\label{tab:example}
\centering
\begin{tabular}{|c|c|c|c|c|c|} \hline
    $\mu$   &   $f$   &   $\Lambda$   &  $n_{s}$   &    $r$   &   $\alpha_{s}$  \\
\hline
    0.01   &   0.01   &   0.00000165   &   0.973024   &   0.188191   &   -0.00482996 \\
\hline
    0.01   &   0.1   &   0.000016   &   0.967254   &   0.177693   &   -0.00350042 \\
\hline
    0.01   &   1   &   0.00018   &   0.960173   &   0.160723   &   -0.00220467 \\
\hline
    0.01   &   5   &   0.0006   &   0.95029   &   0.119164   &   -0.00122968 \\
\hline
    0.01   &   10   &   0.0008   &  0.949265   &   0.0964403   &    -0.00119981  \\
\hline\end{tabular}
\end{table}

\begin{table}[htbp]
\renewcommand{\arraystretch}{1.6}
\caption{Different range of model parameters  $f$,$\mu$ and $\Lambda$ (in units of $M_{P}$)   with Chern-Simon term of the form $\frac{\lambda}{8} \sin\left(\frac{\chi}{f}\right) \epsilon^{\mu \nu \rho \sigma} F_{\mu \nu}^{a} F_{\rho \sigma}^{a}$ when$N=50$:}
\label{tab:example}
\centering
\begin{tabular}{|c|c|c|c|c|c|} \hline
    $\mu$   &   $f$   &   $\Lambda$   &  $n_{s}$   &    $r$   &   $\alpha_{s}$  \\
\hline
    0.01   &   0.01   &   0.00000141   &   0.979907   &   0.153613   &   -0.00343219 \\
\hline
    0.01   &   0.1   &   0.0000141   &   0.979848   &   0.153558   &   -0.00342304 \\
\hline
    0.01   &   1   &   0.000141   &   0.97839   &   0.150695   &   -0.00317821 \\
\hline
    0.01   &   5   &   0.0007   &   0.957516   &   0.102345   &   -0.000391958 \\
\hline
    0.01   &   10   &   0.0008   &  0.958836   &   0.0777509   &    -0.000760981  \\
\hline
\end{tabular}
\end{table}

\begin{table}[htbp]
\renewcommand{\arraystretch}{1.6}
\caption{Different range of model parameters  $f$,$\mu$ and $\Lambda$ (in units of $M_{P}$) with Chern-Simon term of the form $\frac{\lambda}{8} \sin\left(\frac{\chi}{f}\right) \epsilon^{\mu \nu \rho \sigma} F_{\mu \nu}^{a} F_{\rho \sigma}^{a}$ when $N=60$:}
\label{tab:example}
\centering
\begin{tabular}{|c|c|c|c|c|c|} \hline
    $\mu$   &   $f$   &   $\Lambda$   &  $n_{s}$   &    $r$   &   $\alpha_{s}$  \\
\hline
    0.01   &   0.01   &   0.0000012   &   0.979481   &   0.121725   &   -0.00177654 \\
\hline
    0.01   &   0.1   &   0.000012   &   0.979352   &   0.121661   &   -0.00176474 \\
\hline
    0.01   &   1   &   0.00012   &   0.978734   &   0.120114   &   -0.00169073 \\
\hline
    0.01   &   5   &   0.0006   &   0.967059   &   0.0901083   &   -0.000507499 \\
\hline
    0.01   &   10   &   0.0008   &  0.965114   &   0.0652003   &    -0.000516045  \\
\hline
\end{tabular}
\end{table}

From the above tables, we see that the tensor to scalar ratio, $r$ has substantially increased compared to the result reported in the previous section. We also observe that, similar to section II, $r$ increases as $N_e$ decreases. For $N_e = 40$ and $f=0.1$, we find that $r \sim 0.177$ is achieved with $n_s \sim 0.967$. As we go go lower in value of $f$, from $0.1M_P$ to $0.01M_P$, we observe that even if there is an increase in the value of $r$ for lower value of $\Lambda$, $n_s$ goes outside the $5\sigma$ bound of PLANCK . Thus, the generalization of the scalar coupling, from linear to sinusoidal, to Chern-Simon term helps to achieve the large tensor to scalar ratio as predicted by BICEP2.

\section{Discussions:}

In this note we reexamined the Chromo-Natural inflationary model to achieve a large tensor to scalar ratio as predicted by BICEP2. We found that with linear scalar field coupling to Chern-Simon term it is not possible to achieve BICEP2 result.  The model is also not consistent with the combined analysis of BICEP2 and PLANCK if polarized dust power spectra is taken into account in the sense that tensor-to -scalar ratio can not reach the upper bound $0.1$ since this model has a running spectral index.
On the other hand when  the linear coupling is generalized to sinusoidal coupling of scalar field to the Chern-Simon term in the action, we found that the model can successfully achieve large tensor to scalar ratio as predicted by BICEP2 as well as spectral index consistent with PLANCK observations. These results are best obtained for sub-Planckian values of the axion decay constant ($f \sim 0.1 M _P$).

We note that, traditionally, a linear coupling of the field is considered in the Chern-Simon term of the action, possibly a minimal requirement, though the symmetries in this context allows higher order odd powers of the field. Allowing few higher order terms with independent coefficients bring in more number of model parameters which is not a nice thing to do. To avoid this, we choose a simple compact odd function of the field with only one coefficient which is treated as a model parameter. Thus, the model has two important parameters namely the axion decay constant and the said Chern-Simon coupling constants. With these two independent parameters we need to satisfy the observable data on spectral index, tensor-to-scalar ration, running of spectral index besides having number of e-foldings to be order of sixty. Though small values of tensor- to-scalar ratio are possible in the usual Chern-Simon term, it was only for number of e-folding around forty. But in the generalized version, the model can accommodate larger values of tensor- to-scalar ratio.

\underbar{Note Added:}

After our manuscript was put in the arXiv and submitted to the journal for publication, Planck collaboration released the polarization data measuring the polarized dust angular power spectra \cite{RA}. It claims that the dust power is roughly the same magnitude as BICEP2 signal. Further, a quantitative analysis from  both BICEP2 and Planck dust polarization data \cite{CHW} puts the bound on the tensor-to-scalar ratio as $r \le 0.083$ in the $\Lambda$CDM + tensor model. If the model includes running of spectral index this bound can be relaxed to $r \le 0.116$. The model we have considered involves running of spectral index and as we have shown in the last Table, for $N_e =60$, and for both sub-Planckian and Planckian  values of the axion decay constant, the model can accommodate these bounds.

\begin{center}
Acknowledgement
\end{center}
AD and AB thanks Institute of Physics, Bhubaneswar for warm hospitality and infrastructural support. We thank Prof. Ashok Das and Prof J Maharana for valuable discussions.


\end{document}